\documentclass[aps,pre,twocolumn,showpacs,superscriptaddress,preprintnumbers,amsmath,amssymb]{revtex4}
\usepackage{amsmath}
\usepackage{graphicx}% Include figure files
\usepackage{dcolumn}% Align table columns on decimal point
\usepackage{bm}% bold math
\usepackage{float}
\begin{document}

%\preprint{APS/123-QED}
%\begin{CJK*}{GB}{gbsn}  %%%%\begin{CJK*}{GBK}{song}
\title{Parametric study of non-relativistic electrostatic shocks
and the structure of their transition layer}

% Force line breaks with \\

\author{M. E. Dieckmann}\thanks{Electronic mail: Mark.E.Dieckmann@itn.liu.se}
\affiliation{Institute of Physics \& Astronomy, University of Potsdam, 
D-14476 Potsdam, Germany}
\affiliation{Dept of Science and Technology, Link\"oping University,
SE-60174 Norrk\"oping, Sweden}

\author{H. Ahmed}
\affiliation{Centre for Plasma Physics, School of Mathematics and Physics,  
Queen's University of Belfast, Belfast BT7 1NN, United Kingdom}

\author{G. Sarri}
\affiliation{Centre for Plasma Physics, School of Mathematics and Physics,  
Queen's University of Belfast, Belfast BT7 1NN, United Kingdom}

\author{D. Doria}
\affiliation{Centre for Plasma Physics, School of Mathematics and Physics,  
Queen's University of Belfast, Belfast BT7 1NN, United Kingdom}

\author{I. Kourakis}
\affiliation{Centre for Plasma Physics, School of Mathematics and Physics,  
Queen's University of Belfast, Belfast BT7 1NN, United Kingdom}

\author{L. Romagnani}
\affiliation{LULI, Ecole Polytechnique, CNRS, CEA, UPMC, 91128 Palaiseau,
France}

\author{M. Pohl}
\affiliation{Institute of Physics \& Astronomy, University of Potsdam, D-14476 Potsdam, Germany}
\affiliation{DESY, D-15738 Zeuthen, Germany}

\author{M. Borghesi}
\affiliation{Centre for Plasma Physics, School of Mathematics and Physics,  
Queen's University of Belfast, Belfast BT7 1NN, United Kingdom}


\date{\today}% It is always \today, today,

% PACS, the Physics and Astronomy Classification Scheme. %%

\begin{abstract}
Nonrelativistic electrostatic unmagnetized shocks are frequently observed in laboratory 
plasmas and they are likely to exist in astrophysical plasmas. Their maximum speed, 
expressed in units of the ion acoustic speed far upstream of the shock, depends only on 
the electron-to-ion temperature ratio if binary collisions are absent. The formation and 
evolution of such shocks is examined here for a wide range of shock speeds with 
particle-in-cell (PIC) simulations. The initial temperatures of the electrons and the 400 
times heavier ions are equal. Shocks form on electron time scales at Mach numbers between 
1.7 and 2.2. Shocks with Mach numbers up to 2.5 form after tens of inverse ion plasma 
frequencies. The density of the shock-reflected ion beam increases and the number of ions 
crossing the shock thus decreases with an increasing Mach number, causing a slower expansion 
of the downstream region in its rest frame. The interval occupied by this ion beam is on a 
positive potential relative to the far upstream. This potential pre-heats the electrons 
ahead of the shock even in the absence of beam instabilities and decouples the electron 
temperature in the foreshock ahead of the shock from the one in the far upstream plasma. 
The effective Mach number of the shock is reduced by this electron heating. This effect 
can potentially stabilize nonrelativistic electrostatic shocks moving as fast as supernova 
remnant (SNR) shocks.  
\end{abstract}
\maketitle

\section{Introduction}

A supernova explosion accelerates a significant fraction of the material of 
the progenitor star up to a few percent of the light speed c. The initial 
density of the blast shell plasma is many orders of magnitude larger than 
that of the surrounding plasma of the interstellar medium (ISM); the blast 
shell expands freely. The reduction of the plasma density at the front of 
the radially expanding blast shell and its piling up of the ISM plasma 
imply that the latter eventually starts to affect the expansion. A forward 
shock between the upstream ISM plasma and the downstream plasma and a 
reverse shock between the downstream plasma and the (upstream) blast shell 
plasma form close to the blast shell front. The downstream region between 
both shocks expands in time. The shock formation initiates the Sedov-Taylor 
phase of the supernova remnant (SNR) blast shell's expansion \cite{Truelove}, 
in which it can propagate over astronomical distances before it is stopped 
by its interaction with the ISM. 

The low particle collision frequency in the ISM implies that the forward 
shock is collision-less. Its dynamics is dominated by collective plasma 
processes, which imposes constraints on the range of parameters for which 
stable shocks can form. The temperature of the cool ionized component of
the ISM is about 1 eV and the sound speed $c_s \approx 10^4$ m/s. The sound 
speed is defined by $c_s^2 = (\gamma_s k_B / m_i) (T_e + T_i)$ for electrons 
and ions with temperatures $T_e$ and $T_i$ (ion mass $m_i$ and Boltzmann 
constant $k_B$), assuming that the adiabatic constant $\gamma_s$ is equal 
for both species. A fast SNR shock, like the north-eastern (NE) shock of the 
SNR RCW86 \cite{Helder}, would have a Mach number $M_s \sim 10^2-10^3$ in 
the ISM plasma. The weak magnetization of the ISM \cite{Ferriere} suggests that 
SNR shocks are essentially unmagnetized. 

Electrostatic unmagnetized plasma shocks can be generated in laboratory plasmas 
\cite{Honzawa,Romagnani} and in simulation plasmas \cite{ForslundA,ForslundB} 
by the collision of two plasma clouds at an appropriate speed. Collisions between 
identical plasma clouds can result in two types of unmagnetized electrostatic 
shocks \cite{ForslundB}. Sub-critical shocks can convert the entire kinetic 
energy of the inflowing upstream plasma into heat. The shock-reflected ion beam 
of super-critical shocks provides an additional energy dissipation mechanism 
and such shocks are stable at larger Mach numbers than the sub-critical shocks. 
Analytic estimates of the maximum Mach number exist, at which we find stable 
electrostatic shocks.  
The largest estimates of the maximum Mach number of shocks, which develop out
of the collision of two identical plasma clouds, are derived under the 
assumption that the electron temperature $T_{eu}$ exceeds by far the ion 
temperature $T_{iu}$ and that the particle velocity distributions are 
Maxwellian's far upstream of the shock \cite{ForslundA,ForslundB}. The peak
Mach number of subcritical shocks is in this case $M_s \approx 3$, while that
of super-critical shocks is $M_s \approx 6.5$. The peak Mach number of shocks 
decreases with a decreasing ratio $T_{eu} / T_{iu}$ \cite{Bardotti}. No stable 
electrostatic shocks exist under these approximations above a value $M_s 
\approx 6.5$, which is orders of magnitude below those of SNR shocks. 

An asymmetry between the colliding plasma clouds in terms of the electron 
temperature and density yields a double layer \cite{Langmuir} that can raise 
the maximum value of $M_s$ \cite{Sorasio}. Double layers are unstable, because 
the thermal pressure of the dense hot downstream plasma is not necessarily 
balanced by the deceleration or reflection of the upstream plasma. Experimental 
observations indicate that most double layers are rarefactive and some are 
compressive \cite{Raadu} in their rest frame. They may thus not result in stable 
shocks, but they can trigger the formation of stable electrostatic shocks 
\cite{SarriA}, even if the initial Mach number of the flow speed is 100 
\cite{Dieckmann}. 

Shocks can be stabilized for $M_s>6.5$ by a magnetic field. An example is the 
Earth's quasi-perpendicular bow shock \cite{Bale}. It has a Mach number $M_s 
\approx $ 10-12 with respect to the ion acoustic speed, if we take an electron 
temperature of 10 eV of the solar wind and a shock speed of 500 km/s in the rest 
frame of the solar wind. This Mach number exceeds the stability limit of electrostatic 
shocks and the magnetic field must thus provide an additional stabilization. Indeed 
the Earth bow shock is a fast MHD mode shock \cite{Sckopke}. Shocks that propagate into 
an upstream medium with a uniform magnetic field have been examined in the context 
of SNR shocks with particle-in-cell (PIC) simulations 
\cite{LembegeA,Shimada,LembegeB,Lee,Scholer,Chapman,Umeda,Amano,Matsukiyo,Pohl}. 
These studies have addressed their stability, their efficiency in accelerating 
electrons and they have examined wave instabilities in the foreshock \cite{BretA}. 
Usually a relatively strong magnetic field is used due to computational constraints. It is important to determine if a magnetic field is essential for a shock formation and 
for its stable propagation, given that the magnetic field of the ISM is relatively weak.

Unmagnetized electrostatic shocks with a Mach number $M_s = 10^3$ may not exist in 
the ISM. This Mach number is however a least upper bound. It is based on the electron 
temperature of the ISM and neglects any pre-heating of the ISM by the shock. Unless 
the shock is quasi-perpendicular, shock-accelerated electrons or hot downstream 
electrons can escape upstream of the shock. They increase the ion acoustic speed,
which in turn decreases the Mach number. An electron foreshock is observed, for example, 
at the Earth's bow shock \cite{Eastwood}. Hot electrons are detected ahead of this 
shock even though the temperature difference between the electrons in the (downstream) 
magnetosheath and the upstream solar wind is well below that between the downstream 
region of SNR shocks and the ISM. Hotter electrons ahead of an SNR shock result in 
a higher $c_s$ and therefore in a lower value of $M_s$. This pre-heating mechanism 
is independent of the heating of the foreshock by the cosmic ray precursor 
\cite{VolkA,Drury,Vink} and the wave precursors \cite{Ghavamian}. 

The possible existence of electrostatic unmagnetized SNR shocks motivates 
our numerical study. The shock is generated by letting an unmagnetized 
plasma collide head-on with a reflecting wall at the speed $v_c$. This 
corresponds to a collision of two equal plasma clouds at 2$v_c$, because
the particles are reflected elastically. We perform 7 separate simulations 
in which the collision speed between the plasma and the wall is varied from 
1.06$c_s$ to 2.48$c_s$. The temperatures of the co-moving and spatially 
uniform electrons and ions are equal. The downstream plasma and its boundary, 
the shock, expand away from the wall and into the upstream region. The shock 
speed is thus always larger than the collision speed. 

Our results are as follows. All shocks are super-critical and move at a 
speed higher than 1.6 $c_s$ in the upstream reference frame. The simulations 
thus demonstrate that stable shocks can form even if the ions are as hot as 
the electrons, which is not possible according to some analytic models 
\cite{Bardotti}. A higher speed of the inflowing plasma results in a sharper
shock discontinuity and in a larger density of the shock-reflected ion beam 
\cite{ForslundB}. The fraction of ions that can cross the shock decreases and 
the expansion of the downstream plasma is slowed down; doubling $v_c$ from 
1.06 $c_s$ to 2.12 $c_s$ can only raise the shock's Mach number from 1.69 to 
2.31. Shocks form instantly only for $v_c \lesssim$ 1.9 $c_s$. An increase 
of $v_c$ beyond this value delays the shock formation by tens of inverse ion 
plasma frequencies. A collision at $v_c = 2.29 c_s$ results in a shock with 
$M_s = 2.5$ and no shock forms for $v_c = 2.48 c_s$ during the resolved timescale, 
which is comparable to that achieved in laboratory experiments. It may thus not
always be possible to observe fast electrostatic shocks in laser-plasma experiments, while they can develop in the essentially unbounded astrophysical plasmas.

The density of the shock-reflected ion beam, which increases with the collision 
speed, raises the overall ion number density ahead of the shock. This foreshock 
region goes on a positive potential relative to the upstream plasma, which is 
maintained by the ambipolar electrostatic field of the plasma density gradient. 
Upstream electrons are accelerated towards the shock by this electric field. 
The accelerated electrons interact with the electrons, which leak from the 
downstream region into the foreshock. Their mixing raises the mean thermal energy 
of the electrons in the foreshock to about 1/3 of the downstream one for all 
considered cases and the electron velocity distribution becomes a flat-top one 
rather than a Maxwellian one. The simulations suggest that this electron temperature 
depends on the relative speed between the pre-accelerated upstream electrons and 
the leaking downstream electrons and not on the electron temperature far upstream 
of the shock. The implication is that once an electrostatic shock is present it is
no longer appropriate to determine its stability properties using the Mach number
it would have in the far upstream, because it propagates through a much hotter
foreshock plasma.

We outline the particle-in-cell simulation scheme and our initial conditions 
in section 2, we present our results in section 3 and we discuss them in
section 4.

\section{The simulation code and the initial conditions}

\subsection{The model}

\begin{figure}
\includegraphics[width=8.2cm]{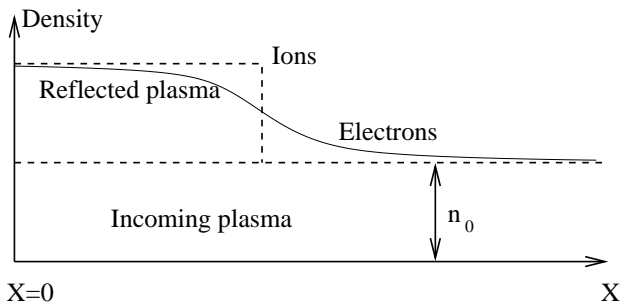}
\caption{The incoming plasma with density $n_0$ moves to the left and 
collides with a reflecting boundary at $x=0$. Initially the density of 
the reflected and incoming plasma adds up to $2n_0$. Electrons stream out 
of the overlap layer due to their larger thermal speed. A charge layer 
builds up at the front of the reflected plasma, which results in an 
electrostatic field. It confines the electrons and accelerates the ions 
to the right.}
\label{fig1}
\end{figure}

The mechanism how a shock forms can be understood with the qualitative model 
shown in Fig. \ref{fig1}. A plasma cloud is reflected at the position $x$=0. 
The cloud is spatially uniform and unmagnetized. It consists of electrons 
and ions, each with the number density $n_0$, that move at the speed modulus 
$v_c$ towards x=0. We assume in this qualitative model that the ions are cold, 
so that the ion front does not spread out in time. The reflected ions and the 
incoming ions interpenetrate and a plasma sheet with an ion density $2n_0$ 
develops close to $x$ = 0. It expands at the speed $v_c$ to larger values of 
$x$. We call this sheet the overlap layer. The higher electron mobility implies 
that some electrons stream out of this layer. They form a thin negatively 
charged sheet just outside of it. An ambipolar electrostatic field is built up 
and the overlap layer goes onto a positive potential with respect to the 
inflowing upstream plasma. The electric field decelerates the ions that enter 
the overlap layer and this plasma compression raises the density of this layer 
above the value $2n_0$. This field also confines the electrons and accelerates 
the surface ions of the overlap layer \cite{SarriA,Sack}. 

If the electric field is strong enough to equalize the speed of all ions 
in the overlap layer, then the ions of both plasma slabs will mix and 
form a single population. This downstream region is bounded by a forward 
shock and it expands due to ion accumulation. The upstream ions and 
electrons are heated up and compressed as they enter the downstream region. 
If all incoming ions enter the downstream region and mix with its plasma, 
then the shock is sub-critical. A fraction of the ions is reflected by a 
super-critical shock. They outrun the shock and form together with the
incoming plasma the foreshock plasma. The upstream region is the one that 
has not yet been reached by shock-reflected ions. 

We employ a 1D simulation geometry, which separates the shock physics from 
that of secondary instabilities and simplifies the interpretation of the 
results. Instabilities can be triggered by the counterstreaming ion beams 
in the foreshock with a relative speed that is larger than the ion acoustic 
speed. However, these instabilities can only develop if wave vectors are 
resolved that are oriented obliquely to the difference vector between the 
mean speeds of both ion beams \cite{ForslundC,Karimabadi,Kato}. Such vectors 
are excluded here by the alignment of the collision velocity vector with the 
simulation direction. The maximum shock speed is a few times the ion acoustic 
speed and thus much smaller than the electron thermal speed, which suppresses 
the Buneman instability that can drive electrostatic waves with a beam-aligned
wave vector \cite{Buneman}.
\subsection{The solved equations and the initial conditions}

A particle-in-cell (PIC) simulation code \cite{Dawson} solves Maxwell's 
equations together with the relativistic Lorentz force equation for an 
ensemble of computational particles (CPs). The collective charge distribution 
$\rho (\mathbf{x})$ and current distribution $\mathbf{J}(\mathbf{x})$ on the 
grid is obtained from the interpolation of the charge and current of all CPs 
onto the grid. The collective charge and current distributions evolve in time 
the electromagnetic fields through the discretized forms of the Amp\'ere's 
and Faraday's laws
\begin{eqnarray}
\frac{\partial \mathbf{E}}{\partial t} = \frac{1}{(\mu_0 \epsilon_0)}\nabla 
\times \mathbf{B} - \frac{1}{\epsilon_0} \mathbf{J},\\
\frac{\partial \mathbf{B}}{\partial t}=-\nabla \times \mathbf{E}.
\end{eqnarray}
They fulfill Gauss' law $\nabla \cdot \mathbf{E} = \rho / \epsilon_0$ and 
$\nabla \cdot \mathbf{B}=0$ either as constraints or through correction steps.
The electric $\mathbf{E}$ and magnetic $\mathbf{B}$ fields update in turn 
the momentum of each CP through the relativistic Lorentz force equation. 
\begin{equation}
\frac{d\mathbf{p}_j}{dt} = q_i \left [ \mathbf{E}(\mathbf{x}_j) + \mathbf{v}_j
\times \mathbf{B}(\mathbf{x}_j) \right ]
\end{equation}
This equation updates the momentum $\mathbf{p}_j$ of the $j^{th}$ particle of 
species $i$. The position is updated as $d\mathbf{x}_j / dt =\mathbf{v}_j$. 
We use the EPOCH PIC code \cite{Brady}.

Our initial conditions are as follows. A plasma consisting of electrons and
ions with mass ratio $m_i / m_e = 400$ is introduced into a one-dimensional 
simulation box with length $L$. We use perfectly reflecting boundary 
conditions at $x=0$ and open boundary conditions at $x=L$. The spatially
uniform plasma fills up the entire simulation box at $t=0$. The particle 
number density of each species is $n_0$ and the initial temperature of 
their Maxwellian velocity distributions is $T_0 = 250 eV$. The electron 
thermal speed $v_e \equiv {(k_B T_0 / m_e)}^{1/2} = 6.625 \times 10^6$ m/s 
and $c_s = 6.05 \times 10^5$ m/s. The Debye length, which includes the
contribution from the ions, is $\lambda_D = v_e / ( \sqrt{2} \omega_p)$, 
where $\omega_p = {(n_0 e^2 / \epsilon_0 m_e)}^{1/2}$ is the electron plasma 
frequency. The ion plasma frequency is $\omega_{pi} = \omega_p / 20$. The 
box length $L$ is resolved by 14400 cells of size $\Delta_x = \lambda_D$. 
We represent the electron species and the ion species by 2000 particles per 
cell, respectively. 

We perform 7 separate simulations with different collision speeds $v_c$:
$v_c = 1.06 c_s$ (Run 1), $v_c = 1.42 c_s$ (Run 2), $v_c = 1.77 c_s$ (Run 3), 
$v_c= 1.95c_s$ (Run 4), $v_c = 2.12 c_s$ (Run 5), $v_c = 2.29 c_s$ (Run 6) 
and $v_c = 2.48 c_s$ (Run 7).

%\vskip .5cm
%\begin{tabular}{|c|c|c|c|}
%\hline
%Run label & $v_c / c_s$ & $M_s$ & $n_r / n_0$ \\
%\hline 
%1 & 1.06 & 1.69 & 1.4 \\
%2 & 1.42 & 1.94 & 1.6 \\
%3 & 1.77 & 2.16 & 1.75 \\
%4 & 1.95 & 2.27 & 1.8 \\
%5 & 2.12 & 2.37 & 1.9 \\
%6 & 2.29 & 2.49 & 2.0 \\
%7 & 2.48 & no shock & \\
%\hline
%\end{tabular}

\section{Simulation results}

Non-relativistic electrostatic shocks are in their most basic form one-dimensional plasma 
structures. We assume that they are planar. Such shocks are fully described by the
distribution of the electrostatic potential along the shock normal $(x)$ direction and by 
the phase space distributions $f_e (x,v_x)$ and $f_i (x,v_x)$ of the electrons and ions, 
respectively. The shock's electrostatic potential $U(x,t) = -\int_0^x E(\tilde{x},t) 
d\tilde{x}$ has to be strong enough to reflect the incoming upstream ions, so we normalize 
it as $2eU/m_i v_c^2$. The potential difference between the far upstream region and the 
downstream region should be $\gtrsim 1$, because the shock speed exceeds $v_c$. Our 
reference potential is $U(x=0,t)$ at the reflecting boundary. We subtract from $2eU(x,t)
/m_i v_c^2$ the offset $2eU_0/m_i v_c^2$, which is the spatio-temporal average of the fully 
developed downstream potential. The offset differs between the simulation runs, because we obtain in
some cases electrostatic ion phase space holes \cite{IonHole} at the wall, which put the wall and the downstream
plasma on different potentials.

The potential $\tilde{U} = 2e ( U(x,t) - U_0 )/m_i v_c^2$ reveals if and when the potential 
difference between the plasma overlap layer and the incoming plasma cloud becomes large enough to reflect the incoming upstream ions. In what follows, we discuss 
separately the results of the simulation runs 1-7 followed by a comparison of the shocks in terms of plasma compression 
and electron heating. 

\begin{figure}
\includegraphics[width=8.2cm]{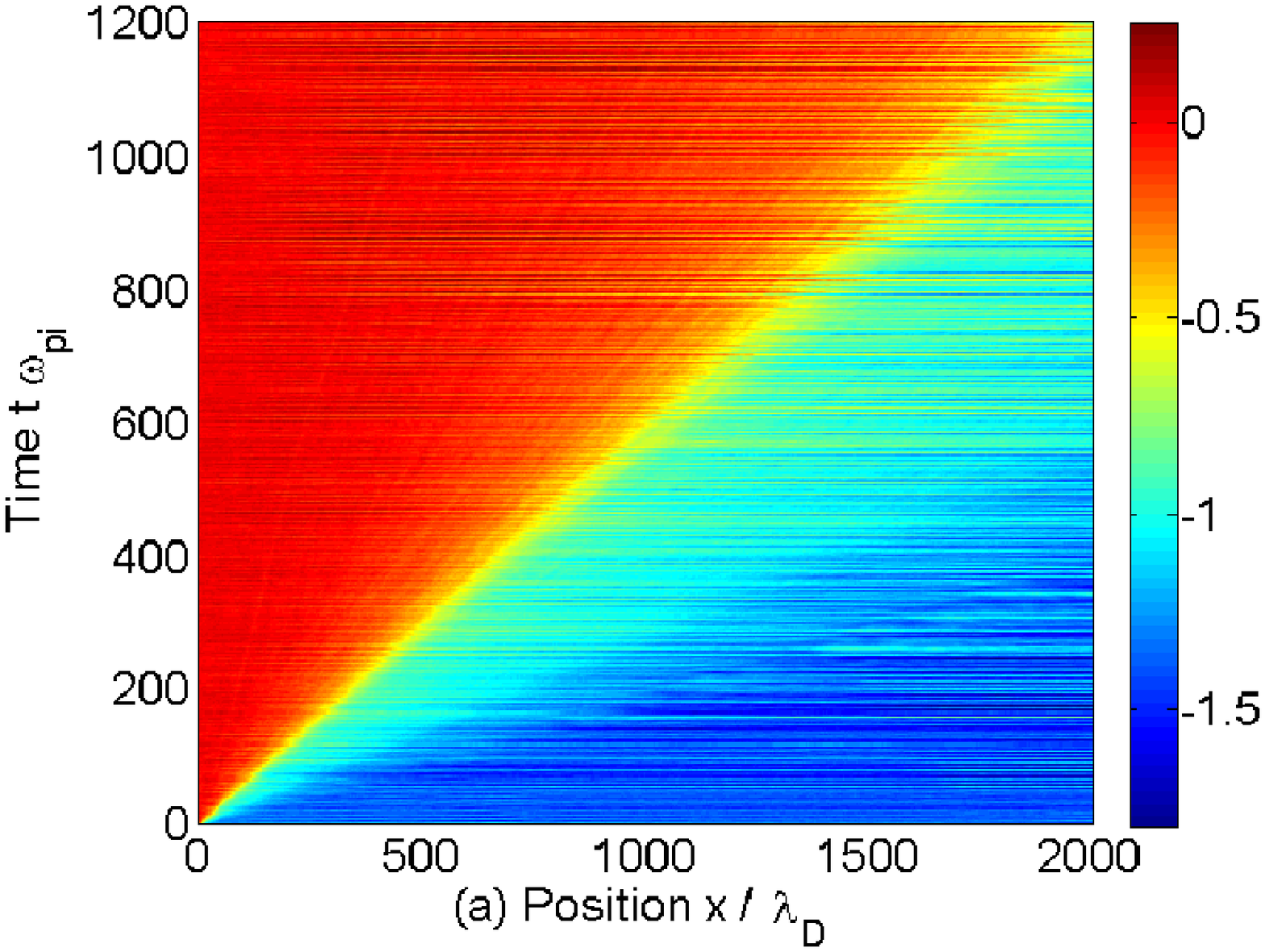}
\includegraphics[width=8.2cm]{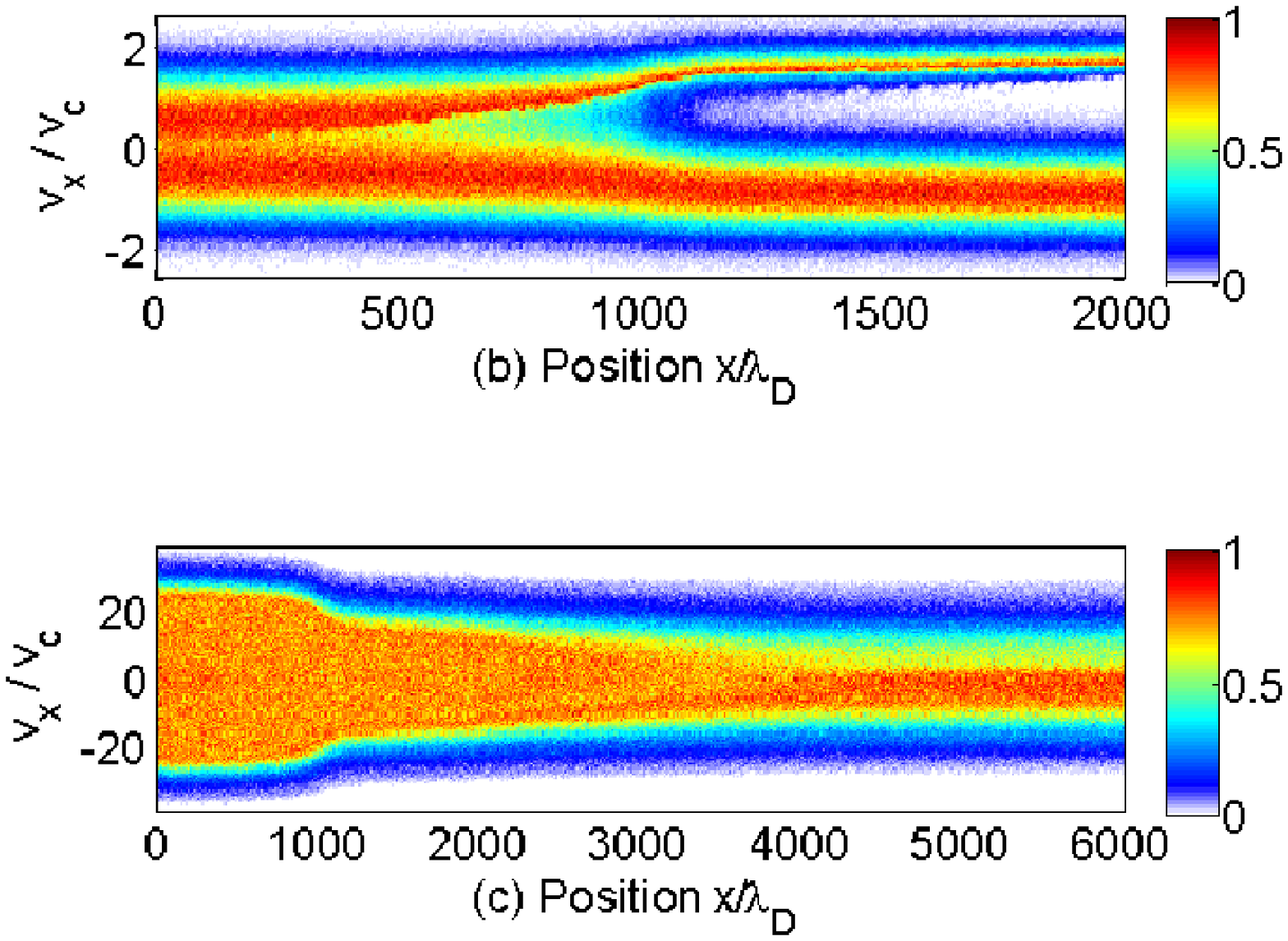}
\caption{(Color online) Collision speed $v_c = 1.06 c_s$ (Run 1): The distribution 
of the normalized electrostatic potential $\tilde{U}(x,t)$ is shown in panel (a). 
The color scale is linear. 
The panels (b) and (c) show the phase space distributions of ions and electrons at 
the time $t\omega_{pi}=600$, respectively. The color scale is linear and normalized 
to the respective peak value.}\label{Run1}
\end{figure}

\textbf{Run 1:} The plasma collides at the speed $v_c = 1.06 c_s$ with the wall and a shock 
should form, which is confirmed by the potential $\tilde{U}(x,t)$ in Fig. \ref{Run1}(a).   
A positive potential forms at the wall on electron time scales. The front of this structure expands 
in the simulation frame at the constant speed $v \approx 2000 \lambda_D / (1200 
\omega_{pi}^{-1})$ or $\approx 0.63 c_s$ for the ion-to-electron mass ratio that we use. 
The front of the potential thus expands at the speed $\approx 1.69 c_s$ in the reference 
frame of the incoming (upstream) plasma. Typical values of the upstream 
potential reach $\approx -1.8$. The potential falls off smoothly in space and 
the width of the transition layer is of the order of $100 \lambda_D$. 

The ion distribution 
in Fig. \ref{Run1}(b) shows that this transition layer is actually wider. The plasma has 
reached a phase space distribution close to $x=0$, which is stable over the considered 
time scale. The velocity distribution is not a single Maxwellian one that is centred at 
$v_x = 0$. We find instead a local minimum of the phase space density at $v_x = 0$ and 
$x\approx 0$ with a value, which is about 20\% below the peak one. The shock can thus 
not fully convert the flow energy of the incoming upstream ions into heat. The width of the 
transition layer of the electrons in Fig. \ref{Run1}(c) is even wider. It expands in 
time because the electrons react to the positive potential of the shock-reflected ion beam. 
The Buneman instability is suppressed by thermal effects and can not account for the observed 
electron temperature rise in the foreshock region. The electrons are heated instead by a 
turbulent mixing of electrons, which are accelerated by the foreshock potential towards the 
shock, with the electrons that leak from the downstream into the foreshock region. The 
downstream electrons (in the interval $x<800 \lambda_D$ at $t\omega_{pi} = 600$) do not have 
a Maxwellian distribution. The distribution $f_e ( x\approx 0, v_x)$ shows a flat top centered 
at $v_x = 0$ and a steep fall-off of the phase space density for $|v_x / v_c| > 20$. A flat-top 
distribution is also observed in the case of the foreshock electrons.

The non-Maxwellian electron distribution alters the shock stability condition \cite{Sharmin} 
and the ions do not get fully thermalized when they cross the shock. These two aspects may 
explain why a shock can form at the high ion temperature and Mach number we consider here, 
which exceeds significantly the limit determined by the analytic model in Ref. \cite{Bardotti}.

\begin{figure}
\includegraphics[width=8.2cm]{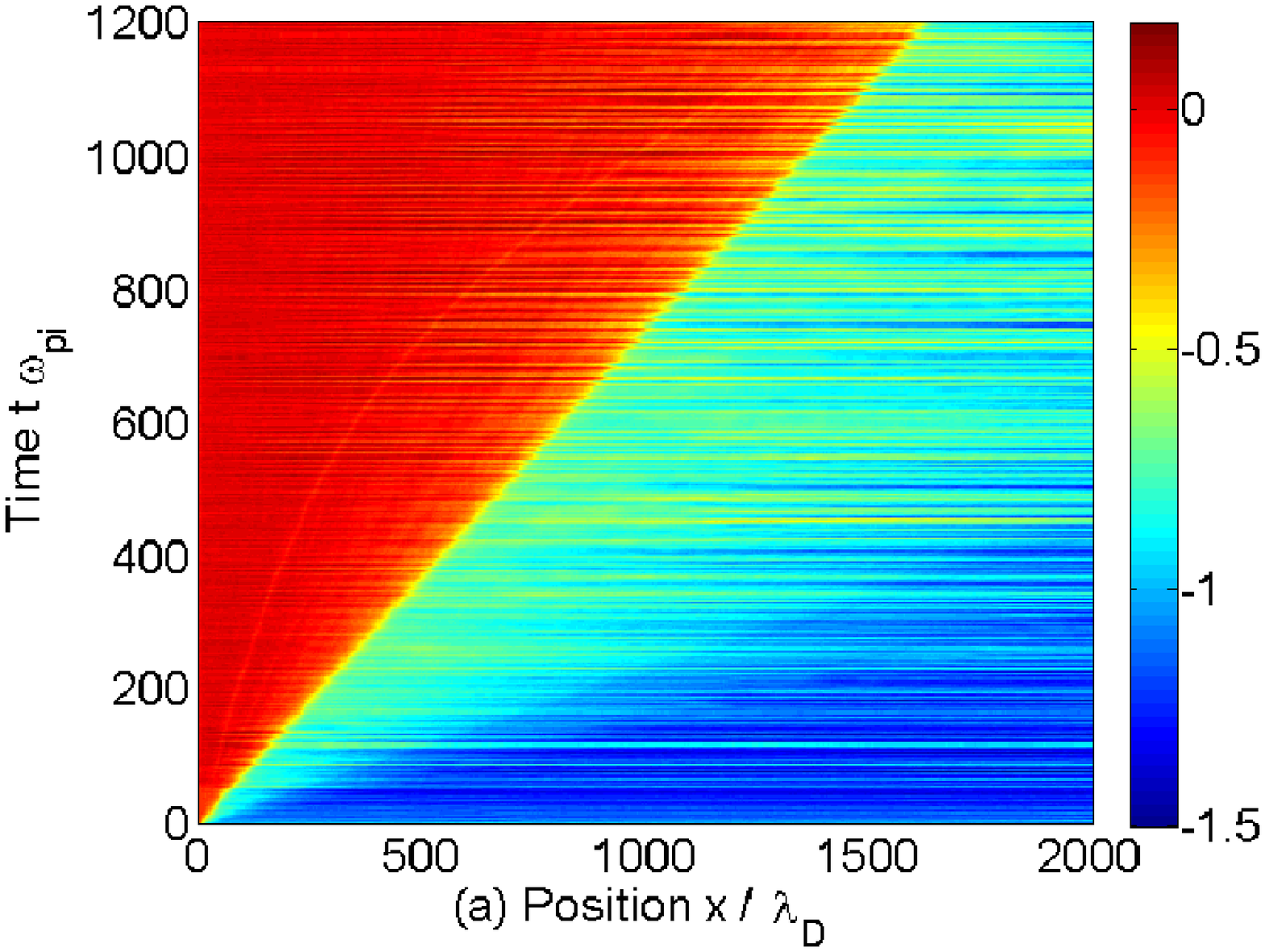}
\includegraphics[width=8.2cm]{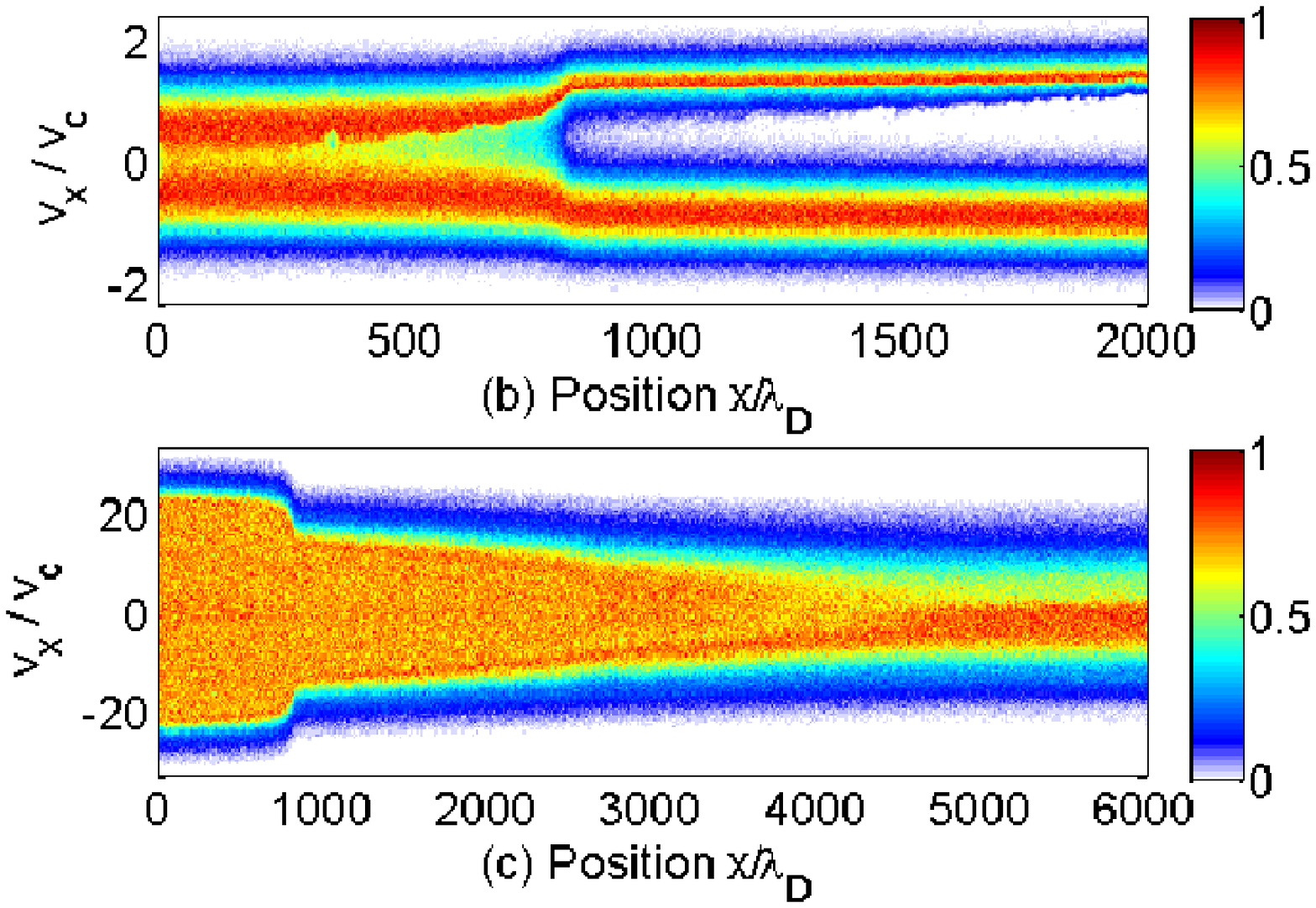}
\caption{(Color online) Collision speed $v_c = 1.42 c_s$ (Run 2): The distribution 
of the normalized electrostatic potential $\tilde{U}(x,t)$ is shown in panel (a). 
The color scale is linear. The panels (b) and (c) show the phase space distributions 
of ions and electrons at the time $t\omega_{pi}=600$, respectively. The color scale is 
linear and normalized to the respective peak value.}\label{Run2}
\end{figure}

\textbf{Run 2:} Raising the collision speed to $v_c = 1.42 c_s$ leaves unchanged the 
qualitative structure of the shock. The potential distribution in Fig. \ref{Run2}(a) shows 
that the shock forms again on electron time scales and that it expands to larger $x$ at a 
constant speed $v\approx 1600 \lambda_D / (1200 \omega_{pi}^{-1})$, which corresponds to 
$0.52 c_s$. This shock thus expands at a lower speed in the simulation frame than the one 
in Run 1. However, the larger $v_c$ implies that its speed in the upstream frame of 
reference is higher with $\approx 1.94 c_s$. 

The ion distribution in Fig. \ref{Run2}(b) shows some qualitative differences compared
to the one in Fig. \ref{Run1}(b). The shock-reflected ion beam appears to be denser, 
which we confirm below. We observe in Fig. \ref{Run2}(b) a pronounced velocity change 
of the incoming upstream ions at $x/ \lambda_D \approx 800$, which we do not find in 
this form in Fig. \ref{Run1}(b). This stronger velocity change should lead according 
to the continuity equation to an increased plasma compression. The electron phase space 
distribution in Fig. \ref{Run2}(c) shows again a flat-top velocity distribution 
downstream of the shock and a broad transition layer towards the upstream region. 

Figure \ref{Run2}(a) shows a weak depletion of the downstream potential that accelerates 
in time. This depletion crosses the position $x/\lambda_D \approx 350$ at the time
$t\omega_{pi} = 600$ and it reaches $x/\lambda_D \approx 1300$ at $t\omega_{pi}=1200$.
This electrostatic field structure is connected to a depletion of the ion phase space 
density at $x / \lambda_D \approx 350$ and $v_x / v_c \approx +0.3$ (Fig. \ref{Run2}(b)). 
We discuss ion phase space holes in more detail below. 

\begin{figure}
\includegraphics[width=8.2cm]{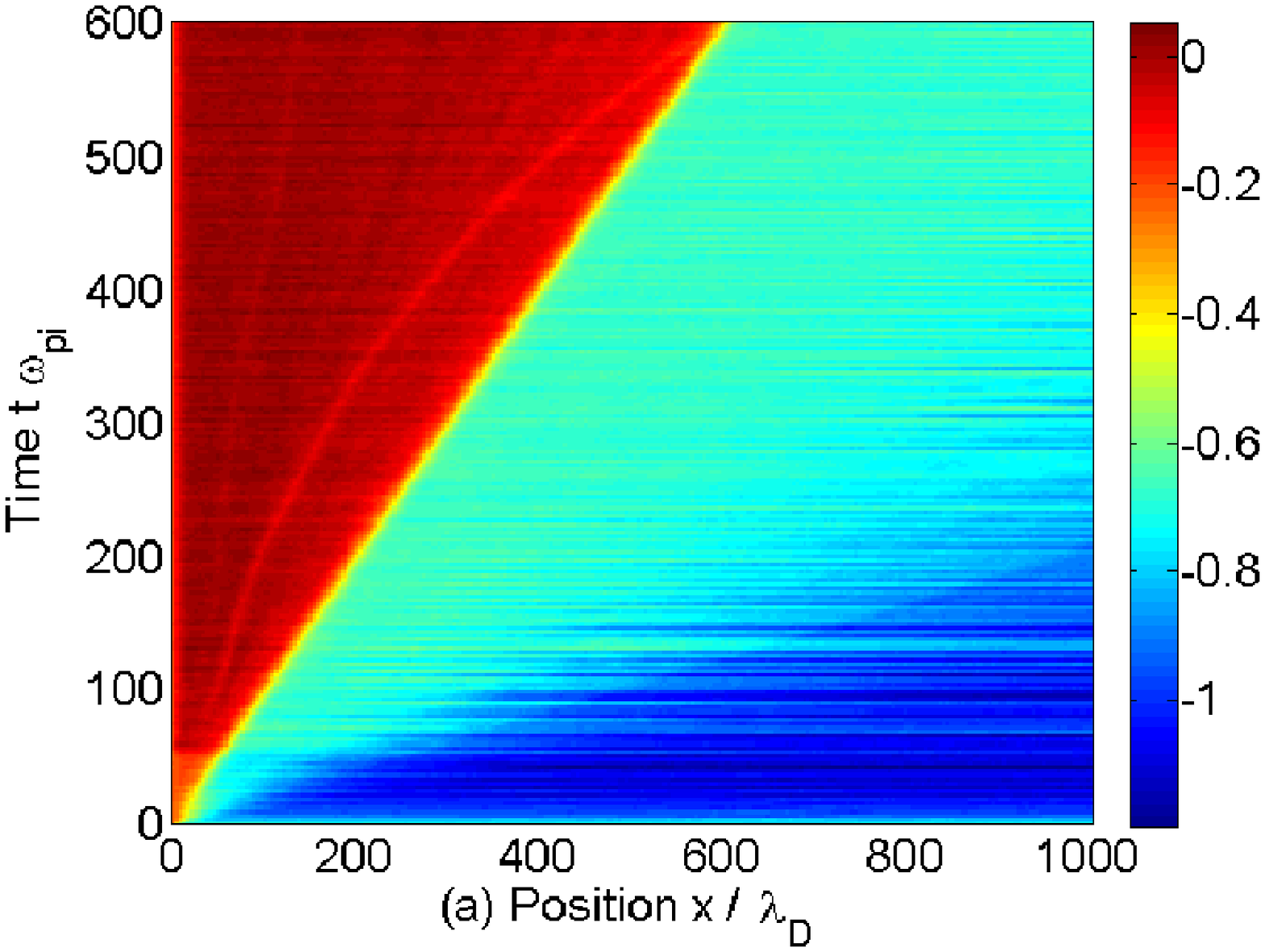}
\includegraphics[width=8.2cm]{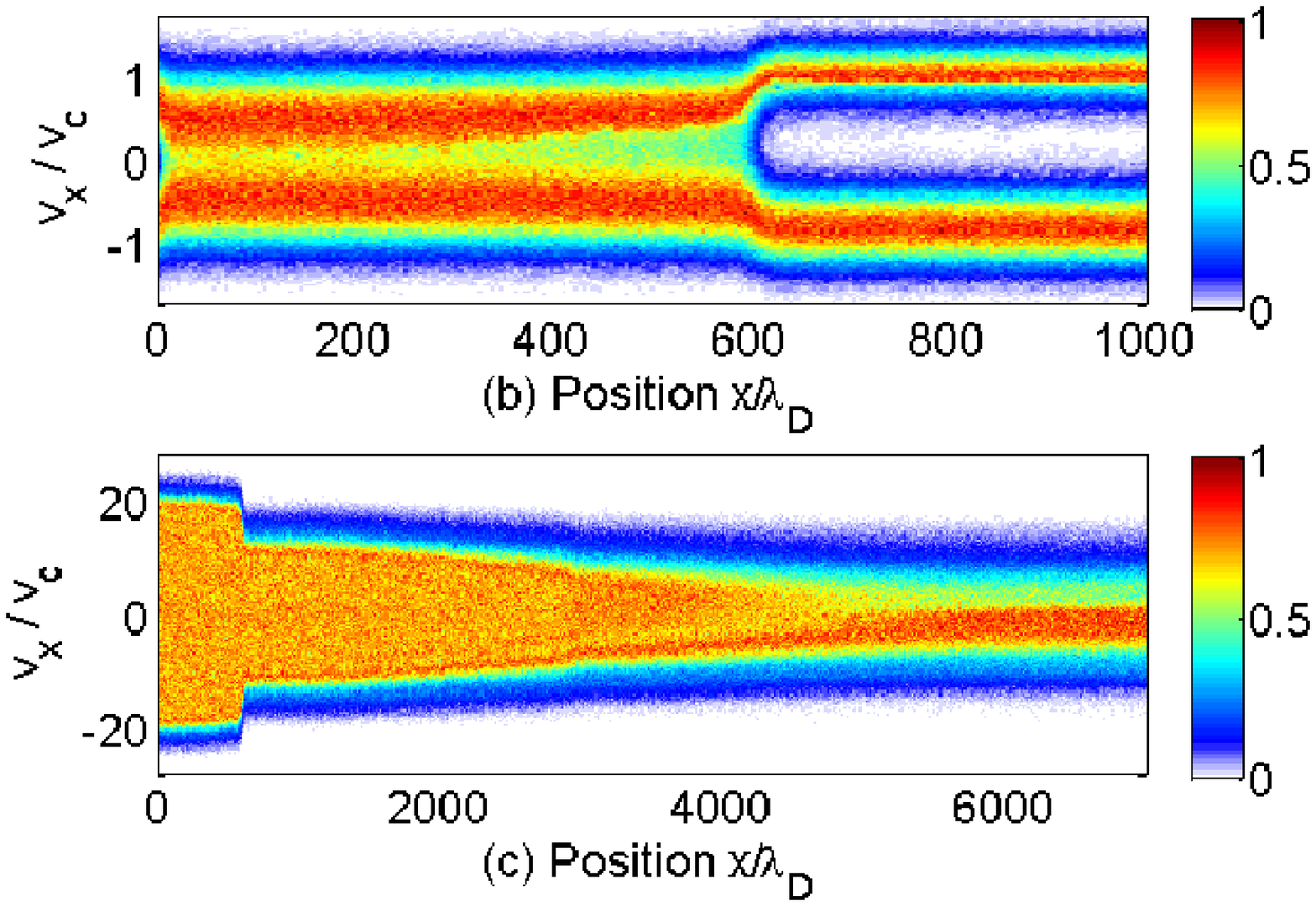}
\caption{(Color online) Collision speed $v_c = 1.77 c_s$ (Run 3): The distribution 
of the normalized electrostatic potential $\tilde{U}(x,t)$ is shown in panel (a). 
The color scale is linear. The panels (b) and (c) show the phase space distributions 
of ions and electrons at the time $t\omega_{pi}=600$, respectively. The color scale 
is linear and normalized to the respective peak value.}\label{Run3}
\end{figure}

\textbf{Run 3:} The collision speed between the plasma and the wall is set to $v_c / 
c_s = 1.77$. The maximum of the positive electrostatic potential in Fig. \ref{Run3}(a) 
still develops close to the wall. The potential jump between the overlap layer and
the upstream region does however not reach the value necessary for a shock formation 
on electron time scales as before. The potential 
jump reaches the value $\Delta \tilde{U} \approx 1$, which is required for a shock 
formation, at the time $t \omega_{pi} \approx 50$. The front of the potential moves 
at the speed $v\approx 600 \lambda_D / (600 \omega_{pi}^{-1})$ or $0.39 c_s$ to larger $x$. 
Its Mach number in the upstream reference frame is $2.16$. Figure \ref{Run3}(a) shows 
a potential structure that outruns the shock. This secondary structure reaches the 
position $x=1000 \lambda_D$ at $t\omega_{pi} \approx 200$. Its speed is comparable to 
$v_c$, which corresponds to the speed at which the reflected ions move to larger $x$. 
The potential jump from $\tilde{U} \approx 0$ to $\tilde{U} \approx -0.7$ at the shock 
front after $t\omega_{pi} = 50$, which separates the downstream region and the shock's 
foot (foreshock), is not enough to reflect ions that move at the speed $-v_c$ towards 
the shock. However, the potential difference between the downstream region and the far 
upstream region is still large enough to reflect ions. 

Figure \ref{Run3}(b) reveals that a shock is present at $x=600 \lambda_D$. This structure 
reflects the incoming ions, which have the lowest speed in the reference frame of the shock, 
and it accelerates those ions from the downstream region, which have the largest speed in 
the direction of the shock. The downstream population of the ions shows again a double-peaked 
velocity distribution that is getting more pronounced as we move away from the wall. This 
separation is upheld by the potential, which changes at $t \omega_{pi} = 600$ from the value 
$\tilde{U} \approx 0$ at $x\approx 200 \lambda_D$ to $\tilde{U} \approx -0.2$ at $x\approx 500 
\lambda_D$. The shock transition layer has an extension comparable to a few hundred $\lambda_D$. The 
electron distribution in Fig. \ref{Run3}(c) shows again the subdivision into downstream 
electrons with $x < 500 \lambda_D$, foreshock electrons in the interval $500 < x / \lambda_D 
< 5000$ and upstream electrons at even larger $x$. 

\begin{figure}
\includegraphics[width=8.2cm]{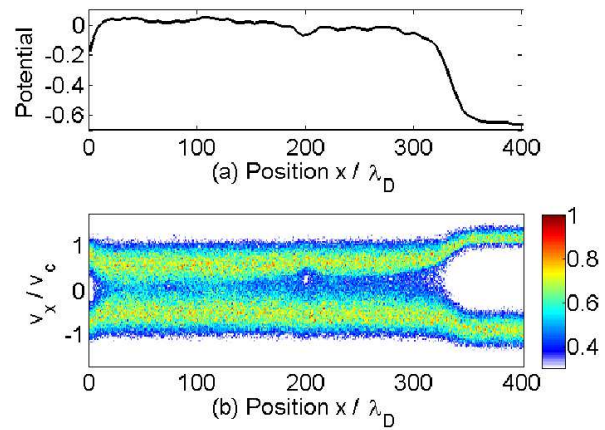}
\caption{(Color online) Panel (a) shows the electrostatic potential close to the
shock at $t\omega_{pi}=300$ (Run 3). Panel (b) displays the associated ion phase space 
density distribution on a linear scale. The potential depression at $x/\lambda_D \approx
200$ in (a) is connected to the ion phase space hole in (b) at the same position and 
centered at $v_x / v_c \approx 0.2$.}\label{Slice}
\end{figure}

The potential distribution in Fig. \ref{Run3}(a) reveals a localized depletion of the 
potential in the downstream region that is similar to the one in Fig. \ref{Run2}(a) but 
stronger. Its curved trajectory implies that it accelerates in time. It reaches the shock 
position at the time $t\omega_{pi} = 600$. This potential depression is caused by an ion 
phase space hole, which is demonstrated by Fig. \ref{Slice}.
Ion phase space holes are structures, in which ions gyrate in a localized depletion of the 
electrostatic potential and form a phase space vortex. This negative potential is upheld 
by a local depletion of the ion charge density \cite{IonHole}. These vortices are tied to 
the ion distribution and move at speeds comparable to the ion thermal speed $v_i$. Its 
apparent acceleration in Fig. \ref{Run3}(a) arises from a change of the bulk speed of the 
ion beam that carries it. 

\begin{figure}
\includegraphics[width=8.2cm]{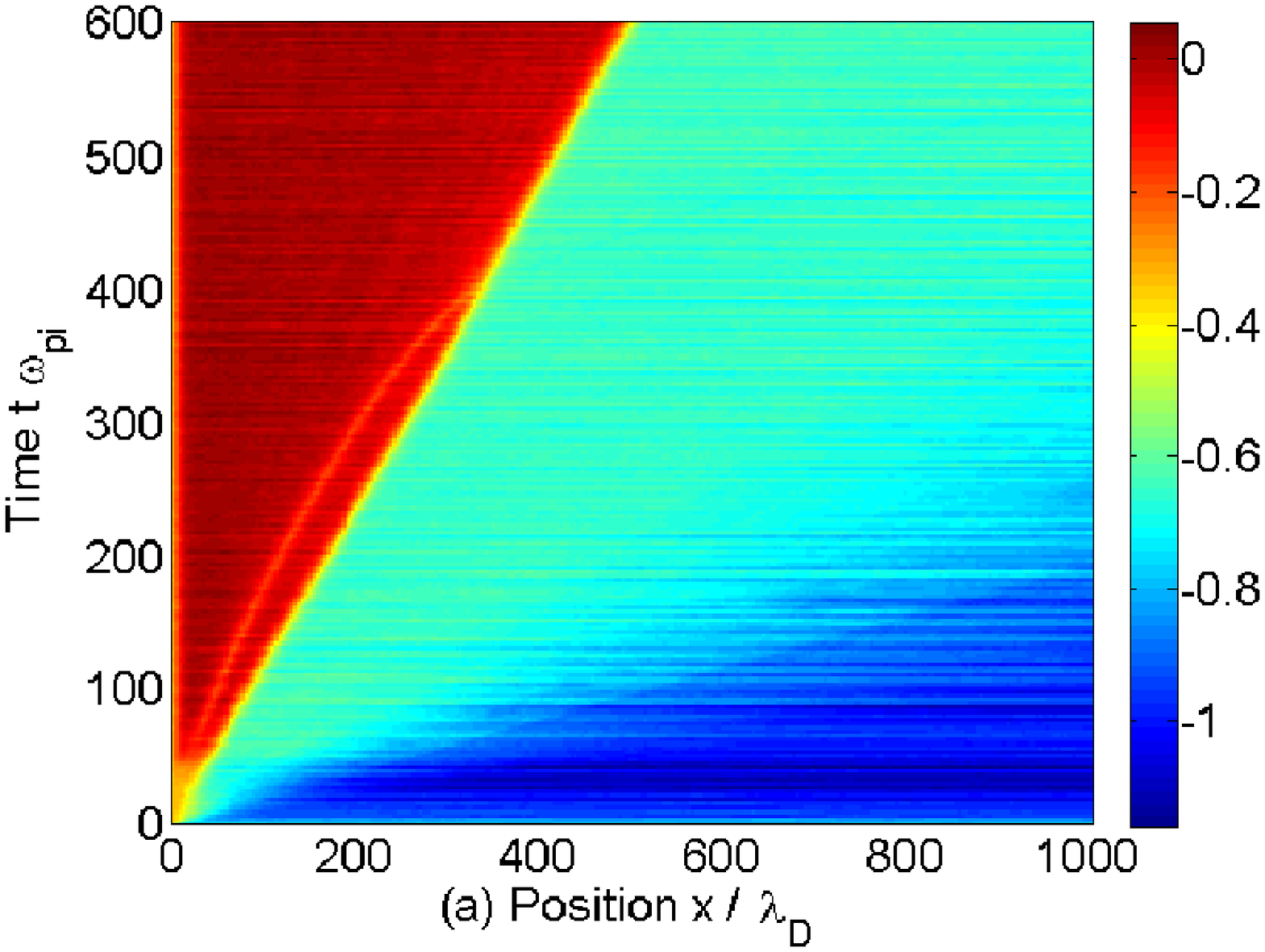}
\includegraphics[width=8.2cm]{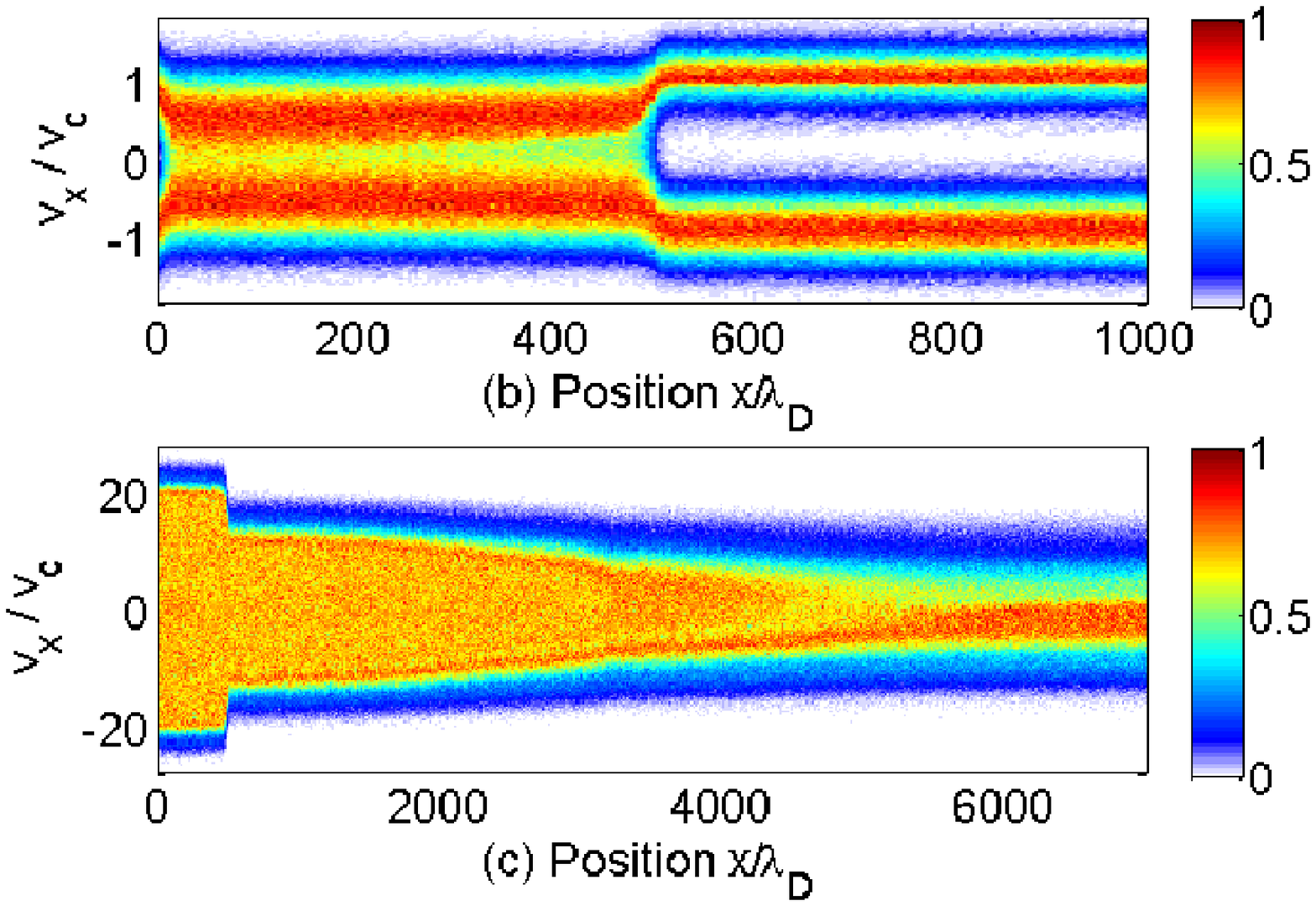}
\caption{(Color online) Collision speed $v_c = 1.95 c_s$ (Run 4): The distribution 
of the normalized electrostatic potential $\tilde{U}(x,t)$ is shown in panel (a). 
The color scale is linear. 
The panels (b) and (c) show the phase space distributions of ions and electrons at 
the time $t\omega_{pi}=600$, respectively. The color scale is linear and normalized 
to the respective peak value.}\label{Run4}
\end{figure}

\textbf{Run 4:} A collision speed $v_c / c_s = 1.95$ results in a shock formation that
is similar to that in Run 3. The potential in Fig. \ref{Run4}(a) demonstrates that
initially only a potential of $\tilde{U}\approx -0.4$ is reached close to the wall. A 
shock forms at $t\omega_{pi} \approx 50$, which propagates at the speed $v \approx 500 
\lambda_D / (600 \omega_{pi}^{-1})$ or $v \approx 0.32 c_s$ to the right, yielding a 
Mach number $M_s \approx 2.27$ in the upstream frame of reference. A shock foot is 
visible that expands at a speed $\approx v_c$ to larger values of $x$. 
The ion distribution in Fig. \ref{Run4}(b) reveals a phase space vortex at $x \approx 0$,
a broad spatial interval $30 < x < 500 \lambda_D$ with an ion distribution that is practically uniform along $x$ and incoming
and reflected ions at $x>500$. The electron distribution in Fig. \ref{Run4}(c) is
qualitatively similar to those in the previous runs. 

\textbf{Run 5:} A simulation with $v_c = 2.12 c_s$ shows a qualitatively different shock
formation stage compared to those in the previous cases. A weak potential $\tilde{U} 
\approx -0.5$ forms first. This structure detaches from the wall and its peak potential 
is located at $x/\lambda_D \approx 40$ at $t \omega_{pi} \approx 70$. This potential 
starts to grow and reaches $\tilde{U} \approx 0$ at $t \omega_{pi} \approx 100$. The potential decreases to $\tilde{U} \approx 
-0.5$ at $t \omega_{pi} \approx 120$. We observe a cyclic reformation of the potential 
with a period of $t\omega_{pi} \approx 50$. Despite this intermittent behaviour, the front 
of the potential propagates at a uniform speed $400 \lambda_D / (600 \omega_{pi}^{-1})$ or 
$0.25 c_s$ into the upstream plasma. The front thus moves upstream at $M_s \approx 2.37$.  
\begin{figure}
\includegraphics[width=8.2cm]{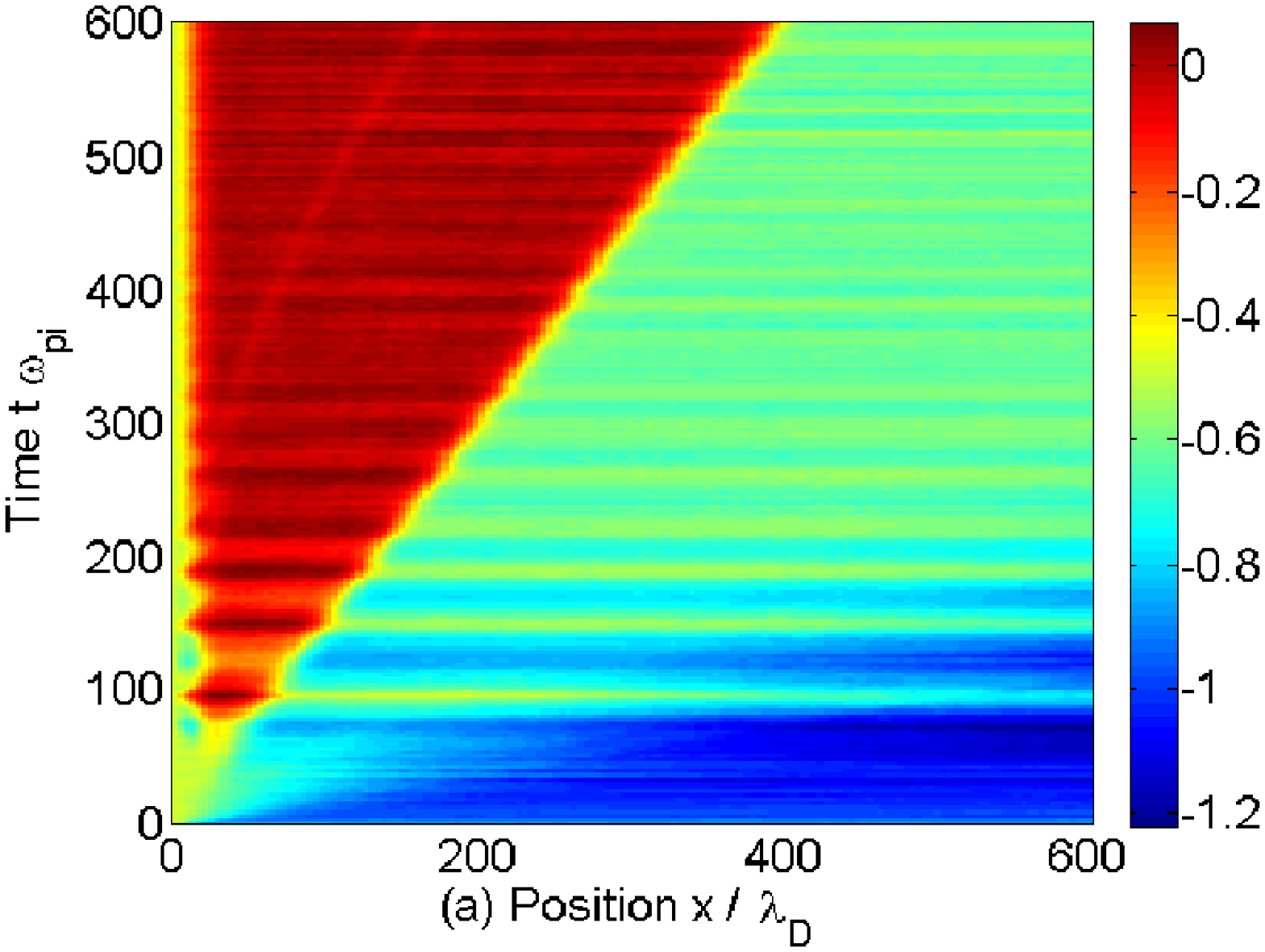}
\includegraphics[width=8.2cm]{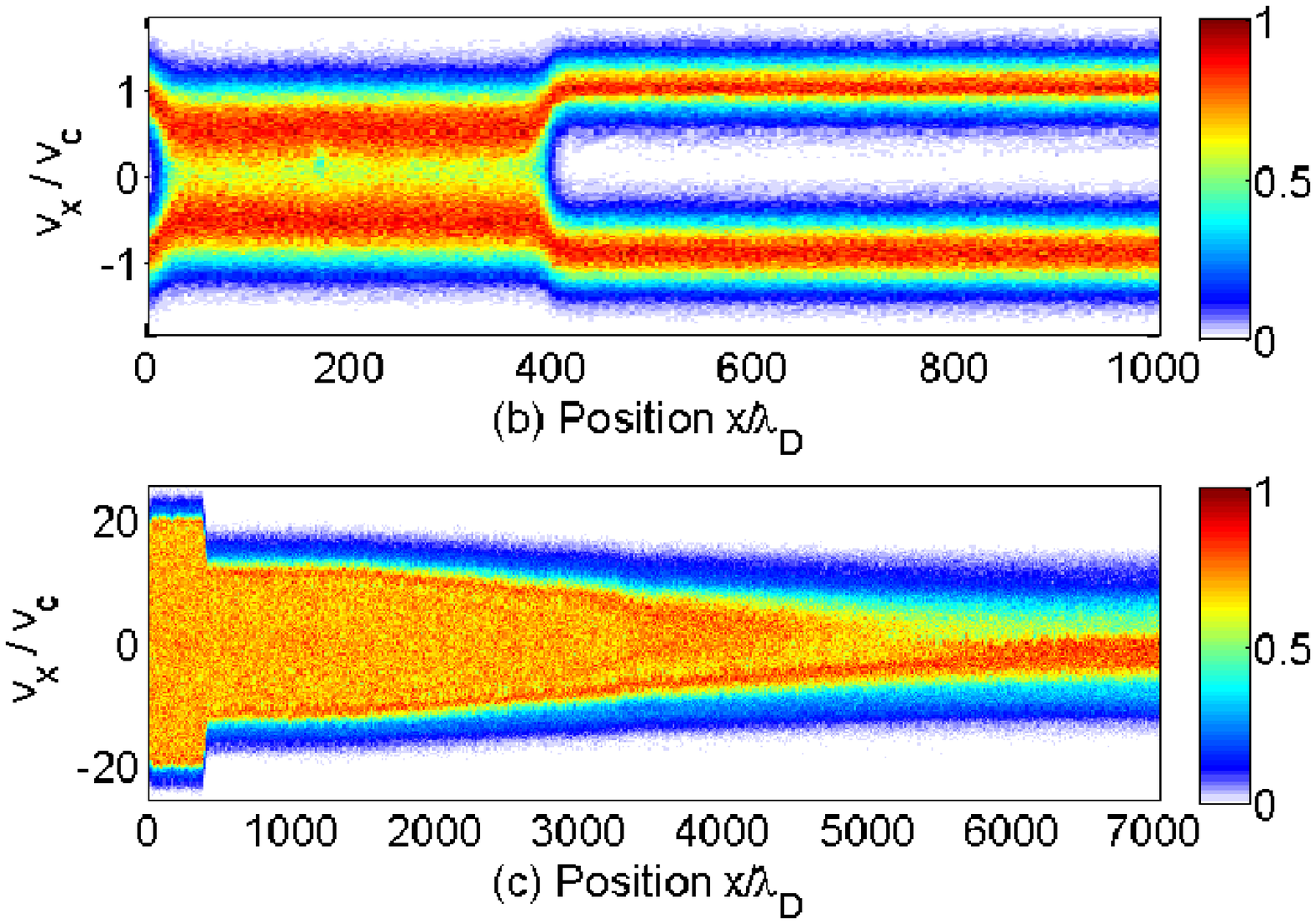}
\caption{(Color online) Collision speed $v_c = 2.12 c_s$ (Run 5): The distribution 
of the normalized electrostatic potential $\tilde{U}(x,t)$ is shown in panel (a). The color 
scale is linear. The panels 
(b) and (c) show the phase space distributions of ions and electrons at the time 
$t\omega_{pi}=600$, respectively. The color scale is linear and normalized to the 
respective peak value.}\label{Run5}
\end{figure}
The particle distributions at $t\omega_{pi} = 600$ in Figs. \ref{Run5}(b,c) agree with 
those observed in the previous two runs, implying that the shock structure does not 
strongly depend on the details of its formation. An ion phase space hole has formed at 
the wall and the electrons have a flat top distribution downstream of the shock and in 
the foreshock.

\begin{figure}
\includegraphics[width=8.2cm]{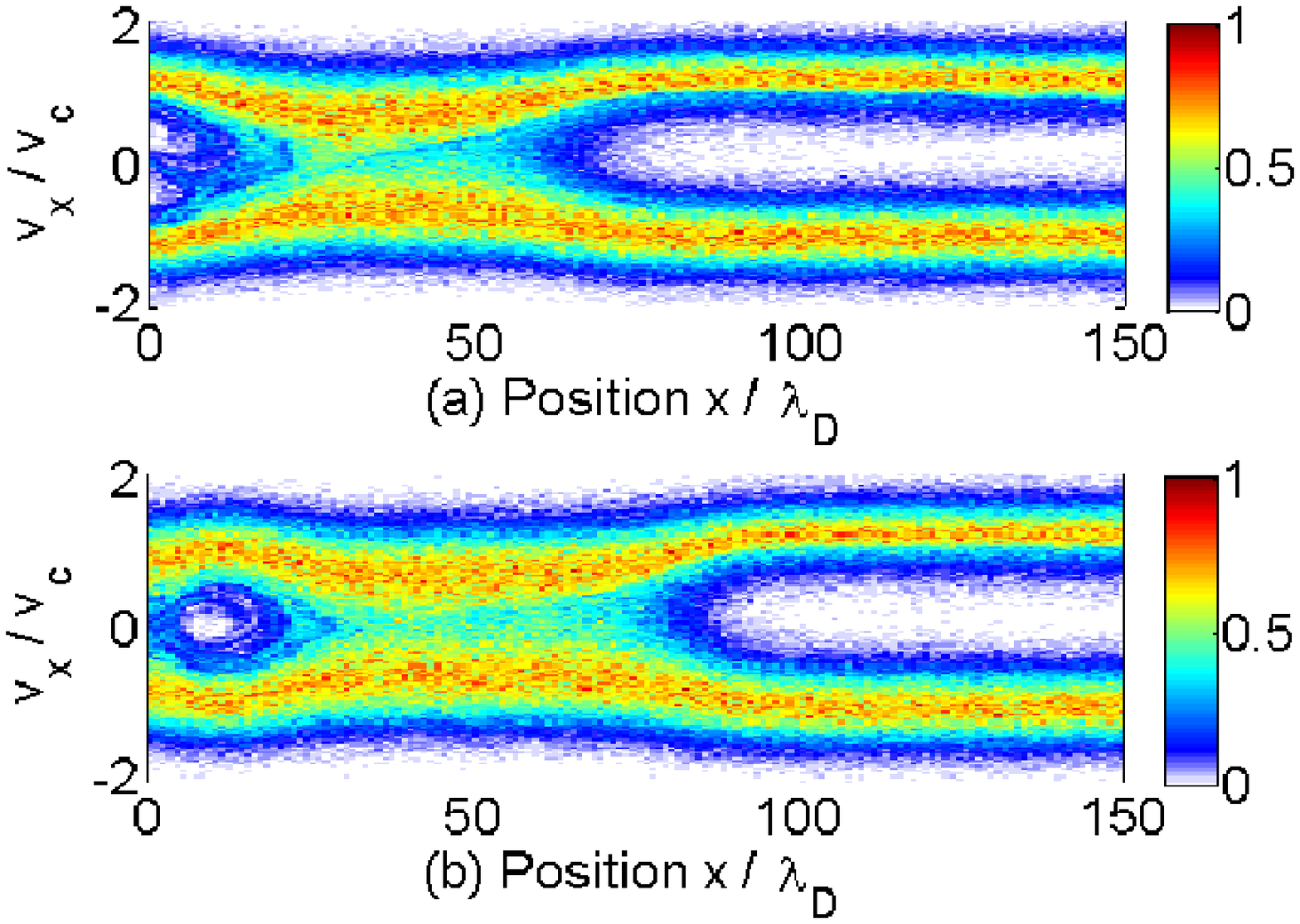}
\caption{Panel (a) and (b) show the ion phase space distribution at the times $t\omega_{pi} 
= 100$ and 120, respectively. Both panels demonstrate that the ion phase space hole at the 
wall with $x<30$ is non-stationary.}\label{Oscillation}
\end{figure}

The cause of the initial oscillations of the potential is a non-stationary ion phase space
hole at the reflecting boundary. Figure \ref{Oscillation} compares the ion phase space 
distributions close to the wall at the times $t\omega_{pi} = 100$ and $t\omega_{pi} = 120$.
The ion phase space hole and thus the minimum of the potential depression are centered at 
$x=0$ at $t\omega_{pi} = 100$. This implies that the plasma overlap layer is on a higher
potential than the wall. The ion phase space hole's center and the minimum of the 
electrostatic potential have moved to $x\approx 10 \lambda_D$ at the later time. The 
difference of the electrostatic potential between the plasma overlap layer and the wall
has been reduced. 

\begin{figure}
\includegraphics[width=8.2cm]{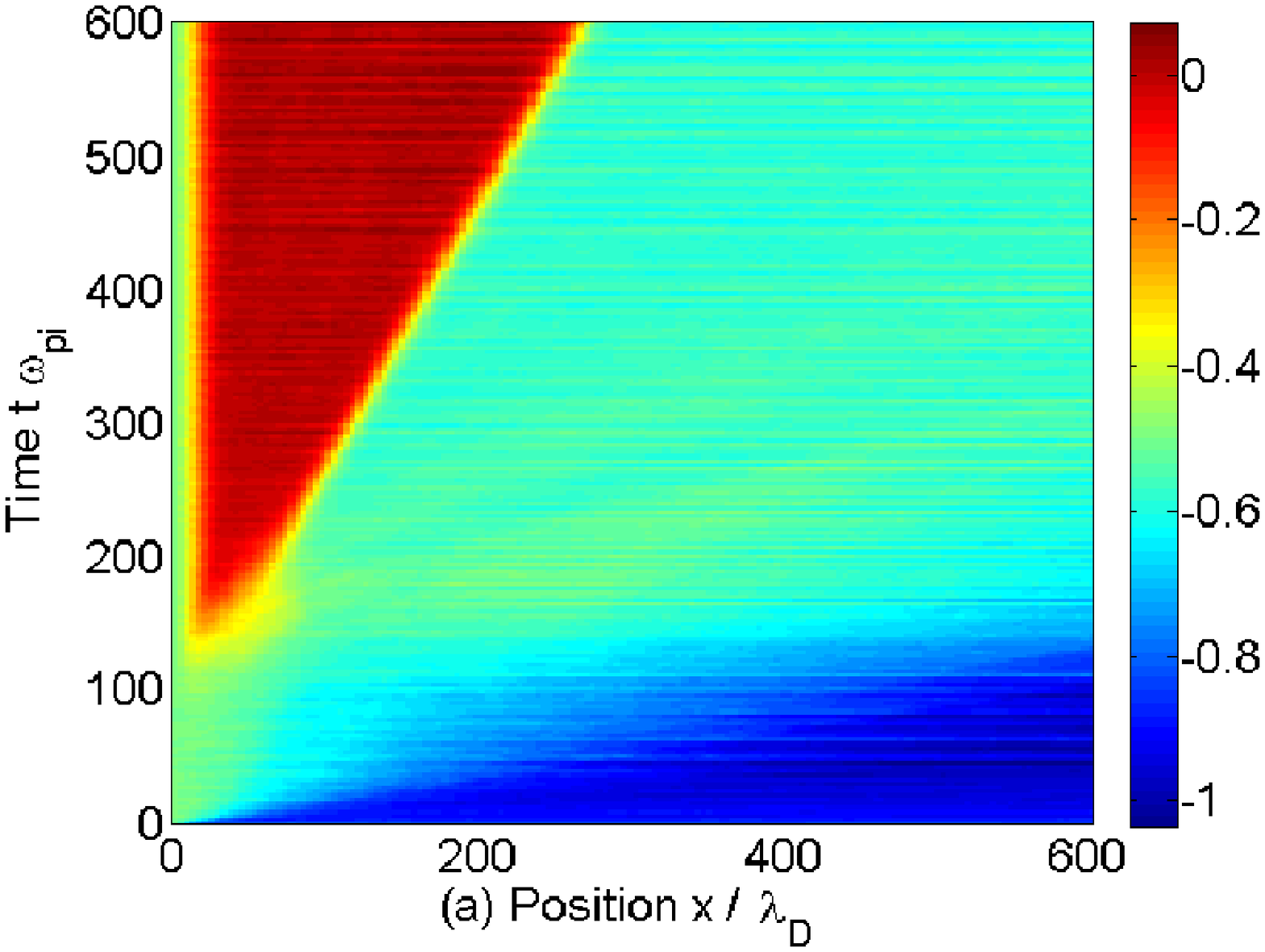}
\includegraphics[width=8.2cm]{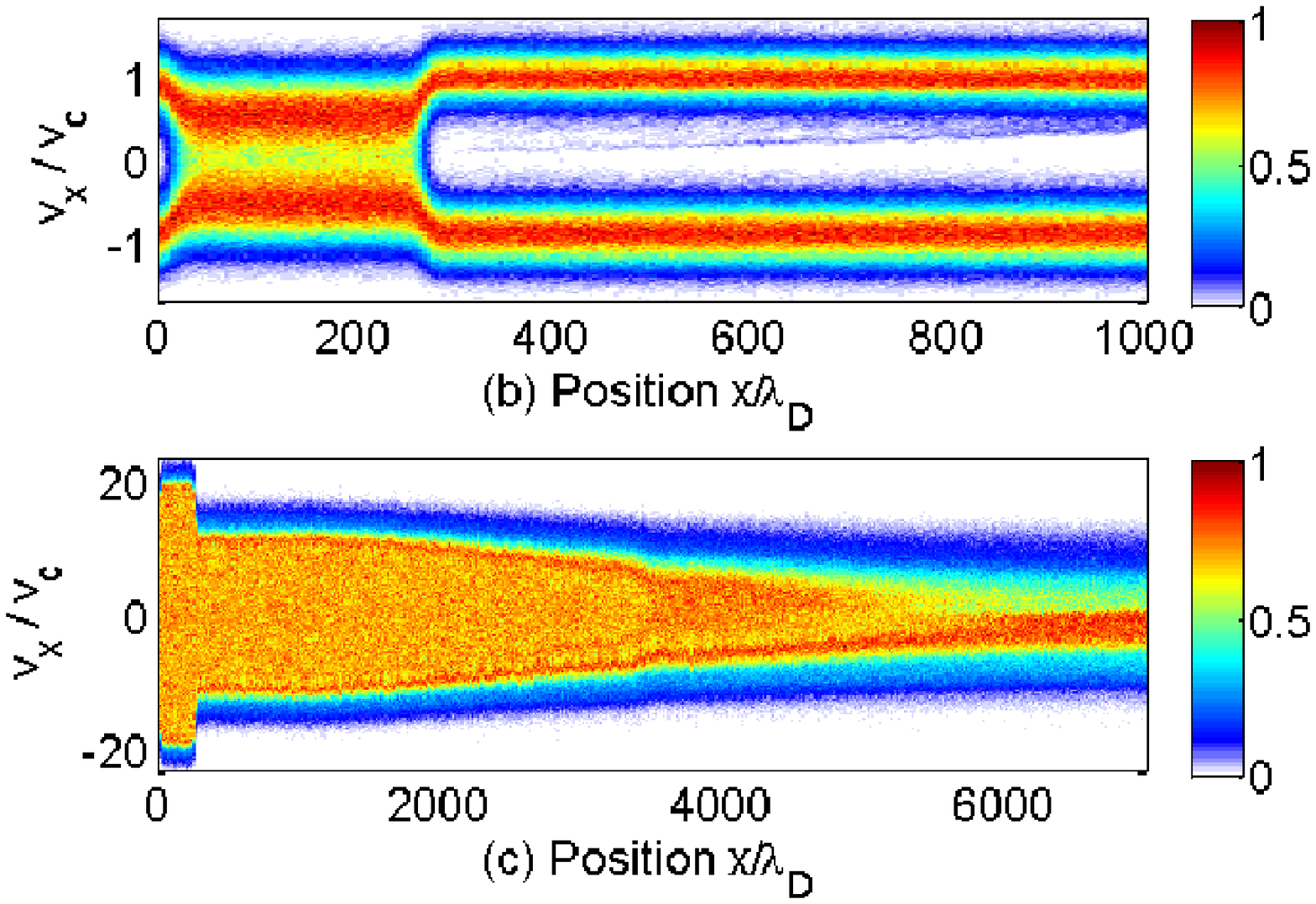}
\caption{(Color online) Collision speed $v_c = 2.29 c_s$ (Run 6): The distribution 
of the normalized electrostatic potential $\tilde{U}(x,t)$ is shown in panel (a). The color 
scale is linear. The panels 
(b) and (c) show the phase space distributions of ions and electrons at the time 
$t\omega_{pi}=600$, respectively. The color scale is linear and normalized to the
respective peak value.}\label{Run6}
\end{figure}

\textbf{Run 6:} A collision speed $v_c = 2.29 c_s$ is the largest one that results
in a shock during a simulation time $t\omega_{pi} = 600$. The shock does not form
instantly, in line with the results of the simulation runs 3-5. It forms close to the wall and
the potential distribution in Fig. \ref{Run6}(a) shows no intermittent behaviour.
The latter is thus not a consequence of large values of $v_c$. A strong and steady 
foreshock potential is observed, which is again tied to the shock-reflected ion beam. 
We notice that a straight line fit of the shock front does not intersect $x=0$ at $t=0$ and the shock has thus not formed at the front of the plasma cloud. 
The ion distribution in Fig. \ref{Run6}(b) agrees qualitatively with those in the
simulation runs 3-5 with respect to the subdivision into the ion phase space hole at the wall,
a downstream region with a double-peaked velocity distribution and the foreshock
structure consisting of incoming and reflected ion beams. 

\begin{figure}
\includegraphics[width=8.2cm]{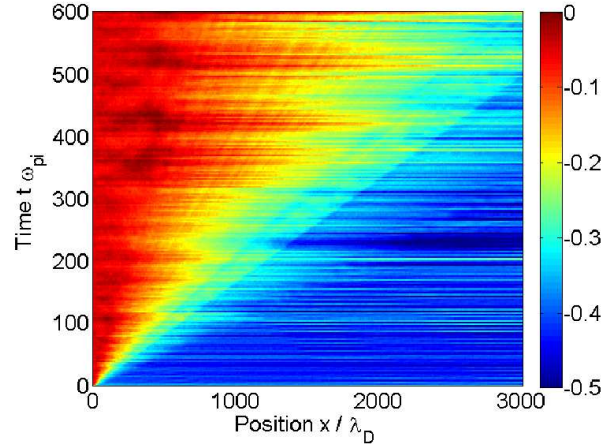}
\caption{(Color online) The distribution of the normalized electrostatic potential 
$\tilde{U}(x,t)$ computed from a simulation with $v_c = 2.48 c_s$ is shown. The color 
scale is linear.}\label{Run7}
\end{figure}

An increase of the collision speed to $v_c = 2.48 c_s$ does not result in a shock
formation during $t\omega_{pi} = 600$, which is demonstrated by the corresponding
$\tilde{U}(x,t)$ displayed in Fig. \ref{Run7}.
The potential difference between the plasma overlap layer and the far upstream
region is not sufficient to yield a shock; no localized strong jump of the
electrostatic potential develops that could be associated with a shock. The ion
phase space distribution (not shown) does not show a shock either.

A comparison of the ion distributions in the simulation runs 1-6 shows two 
trends. The 
first trend is a shrinking of the ionic shock transition layer that connects 
the downstream region with its spatially uniform double-peaked velocity 
distribution and the upstream region that consists of incoming and reflected 
ions. The width of this interval, which is characterized by maxima in the ion 
velocity distribution that diverge with increasing values of $x$, is $500 
\lambda_D$ in Fig. \ref{Run1}(b) and it shrinks with increasing $v_c$ to reach
a few tens of $\lambda_D$ for the highest collision speed. The second trend, 
which has already been reported in Ref. \cite{ForslundB}, is that the density 
of the shock-reflected ion beam increases. The velocity distribution of the 
incoming ions and the reflected ions is practically symmetric with respect to 
$v_x=0$ for $x>300 \lambda_D$ in Fig. \ref{Run6}(b). The shock thus reflects 
practically all incoming ions and the low number density of ions that make it 
downstream explains the low expansion speed of $250 \lambda_D / (500 
\omega_{pi}^{-1})$ or $0.2c_s$ in Fig. \ref{Run6}(a). Its Mach number is 
$M_s \approx 2.5$. 

A comparison of the electron phase space distributions obtained from the 
simulation runs 
1-6 also reveals a trend. The distributions show a striking similarity if the 
velocity axis is scaled by $v_c$. An important feature is that the thermal 
spread of the electrons in the foreshock is practically unchanged by the 
choice of $v_c$. We attribute this to the way they are heated up. Figures 
\ref{Run1}(c)-\ref{Run3}(c) and Figs. \ref{Run4}(c), \ref{Run5}(c),
\ref{Run6}(c) show that the electron's phase space distribution 
is bounded by two populations. The electrons with the largest positive values 
of $v_c$ are supplied by downstream electrons that leak through the shock. We 
can determine the electron population that can leak into the foreshock as 
follows. The shock potential in the case of Run 6 can reflect ions that move 
from the upstream region towards the shock at the speed $v_c$. The potential 
can thus confine downstream electrons with a velocity $\lesssim {(m_i / 
m_e)}^{1/2}v_c$. The fastest downstream electrons in Fig. \ref{Run6}(c) have 
a speed modulus $\approx 25 v_c$ and they can not be confined by the shock 
potential. They have lost a significant fraction of their kinetic energy by 
the time they have travelled far upstream but they are still fast in the 
foreshock region. The foreshock electrons with the largest negative speeds 
are supplied by the upstream electrons, which are accelerated towards the 
shock by the positive potential of the foreshock. Both counter-streaming 
electron populations thermalize in the foreshock. 

In what follows we want to provide further support for the trends outlined above. We 
analyse for this purpose the normalized ion density $n_i (x) = n_0^{-1} \int f_i (x,v_x) 
dv_x $ and the mean kinetic energy per electron $\tilde{T}_e (x) = n_e (x{)}^{-1} \int 
v_x^2 f_e (x, v_x) dv_x $ with $n_e (x) = \int f_e (x,v_x) dv_x$. We normalize $T_e (x)$ 
by the mean kinetic energy per electron of a Maxwellian velocity distribution with the temperature $T_0$. 

\begin{figure}
\includegraphics[width=8.2cm]{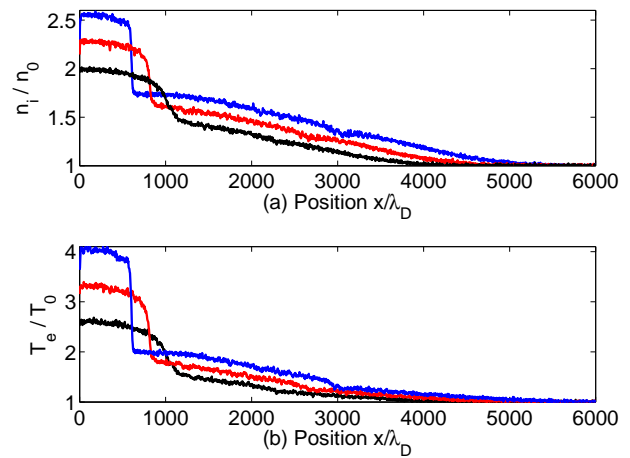}
\caption{(Color online) The total ion density in units of $n_0$ is shown in 
panel (a). The mean thermal energy per electron normalized to that for $T_0 
= 250$ eV is shown in panel (b). The lowest (black) values corresponds to 
$v_c / c_s = 1.06$, the intermediate (red) values to $v_c / c_s = 1.42$ and
the largest (blue) values to $v_c / c_s = 1.77$. The time is $t\omega_{pi}=600$.}
\label{DensityTemp1}
\end{figure}

\begin{figure}
\includegraphics[width=8.2cm]{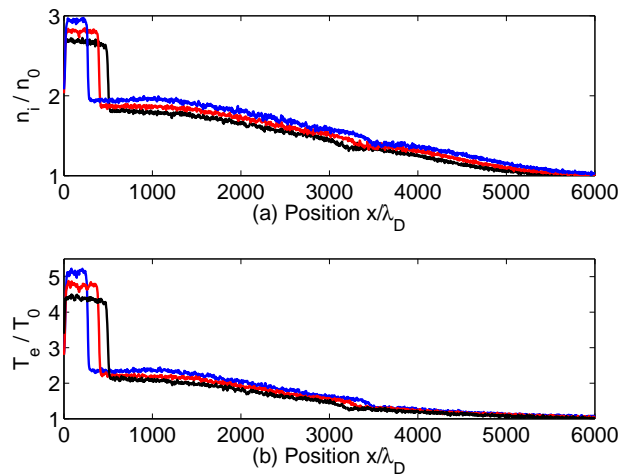}
\caption{(Color online)  The total ion density in units of $n_0$ is shown in 
panel (a). The mean thermal energy per electron normalized to that for $T_0 
= 250$ eV is shown in panel (b). The lowest (black) values corresponds to 
$v_c / c_s = 1.92$, the intermediate (red) values to $v_c / c_s = 2.12$ and
the largest (blue) values to $v_c / c_s = 2.29$. The time is $t\omega_{pi}=600$.}
\label{DensityTemp2}
\end{figure}

We compare the ion density distributions in the Figs. 
\ref{DensityTemp1}(a) for run 1-3 and \ref{DensityTemp2}(a) for run 4-6. The corresponding 
mean thermal energies of the electrons are compared in the Figs. \ref{DensityTemp1}(b) and 
\ref{DensityTemp2}(b). All curves are measured at the time $t\omega_{pi} = 600$. We capture 
the full spatial interval of width $6000 \lambda_D$ affected by the shock-reflected ions. 

The ion densities downstream and, thus, the plasma compression increase with an increasing 
value of $M_s$. This is a well-known fact that is here confirmed by the simulations. A 
somewhat surprising observation is that the downstream density in run 1 is only $2n_0$, 
which corresponds to a mere superposition of the density contributions of the incoming 
and reflected ions. The electrostatic potential in Fig. \ref{Run1}(a) was not strong
enough to slow down significantly the incoming ions (See Fig. \ref{Run1}(b)) and the
ions are thus not strongly compressed.

The curves for $n_i (x)$ demonstrate that the width of the shock transition layer shrinks 
with an increasing value of $M_s$. The distribution $n_i(x)$ for run 1 changes from the 
downstream value to the foreshock value over several hundreds of $\lambda_D$, while the 
ion density is practically discontinuous on the displayed spatial scale in run 6. The ion density 
in the shock's foot increases with $M_s$. It is $\approx 1.5 n_0$ immediately ahead of the 
shock in run 1 and it gradually increases up to $\approx 2n_0$ in run 6. The latter value 
implies that most ions are reflected by the shock; the shock acts as a piston \cite{ForslundB}. 

The curves for the mean thermal energy per electron follow the trend of the ion
densities. The mean electron thermal energy is only slightly elevated in the
foreshock region in run 1. The foreshock electrons get hotter with increasing
$v_c$ and the typical thermal energy of electrons in the foreshock has more than 
doubled in run 6 compared to the value far upstream. The shock in run 6 thus moves
through a plasma with an elevated sound speed since $c_s \propto {(T_e + T_i)}^{1/2}$. This 
temperature rise is not large in our simulations. It can nevertheless be important
because it is caused by the mixing of leaking downstream electrons that move upstream 
and upstream electrons that have been accelerated towards the shock by the foreshock 
potential. The energy available to the incoming upstream electrons is dominated by
the bulk kinetic energy they gain in the foreshock potential and not by their thermal 
energy. The foreshock's electron temperature should thus be robust against changes 
of the electron temperature far upstream. The mean thermal energy of the electrons 
converges with increasing values of $x$ to the initial value and the rise in the electron 
temperature is thus limited to the foreshock region.

\begin{figure}
\includegraphics[width=8.2cm]{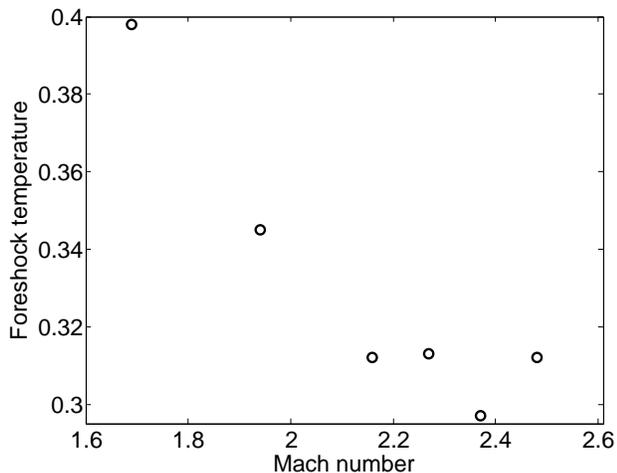}
\caption{The mean thermal energy of the foreshock electrons in units of the downstream value.}
\label{TempRatio}
\end{figure}

The mean thermal energies of the foreshock electrons and the downstream electrons 
increase with $v_c$. We can compute from Figs. \ref{DensityTemp1}(b) and 
\ref{DensityTemp2}(b) the ratio of the mean electron thermal energies in the foreshock 
and downstream. We compare for this purpose the respective values at the positions 
$x=150 \lambda_D$ and $x=1500 \lambda_D$ in Fig. \ref{DensityTemp1}(b) and those at 
$x=100 \lambda_D$ and $x=1000 \lambda_D$ in Fig. \ref{DensityTemp2}(b). The measured 
ratios are displayed in Fig. \ref{TempRatio} as a function of the shock's Mach number.
We find that the temperature of the foreshock electrons is in all cases about 1/3 of
the value in the downstream region.

\section{Discussion}

We have performed here a series of one-dimensional particle-in-cell (PIC) simulations
of electrostatic shocks. The purpose has been to gain insight into the conditions under
which such shocks form and into the structure of their transition layer for a wide range 
of shock speeds. These aspects are important in the context of supernova remnant (SNR) 
shocks. Their large nonrelativistic speeds exceed the ion acoustic speed in the ISM by a 
factor $10^2-10^3$. This factor is much larger than the limit $\lesssim 6.5$ for stable shocks, which 
is derived from various theoretical models \cite{ForslundA,ForslundB,Bardotti}. It is 
thus essential to know the degree to which the ISM's temperature is important for SNR 
shocks. In other words, can a shock, once it is present, generate the electron temperatures 
it needs to be stable? This parametric study is also relevant for laboratory experiments. 
Important questions are what kind of shocks can form during the limited time over which 
observations are possible and what properties these shocks have. 

Our study addressed
shocks in initially unmagnetized plasma that develop out of the collision of two equal
plasma clouds at a speed that is less than 1\% of the light speed $c$. Such shocks
remain electrostatic because the low flow speeds yield only weak currents and magnetic fields. A speed in
excess of $0.3c$ results in the growth of magnetic fields \cite{Kazimura,Kato,Pohl} that 
are strong enough to affect the shock dynamics and they become dominant at ultrarelativistic 
speeds even if the plasma was initially unmagnetized \cite{Spitkovsky}. The chosen collision
speeds are realistic for the late expansion phase of SNR blast shells. Our parametric study 
has provided the following results. 
  
Shocks form for the considered electron and ion temperatures up to a Mach number $\approx$ 
2.5. A shock with this Mach number reflects practically all incoming ions and acts
as a piston. Raising the plasma collision speed by another 8\% results in a shock-less 
interpenetration of the plasma clouds. Slower shocks with a Mach number $M_s \le 1.9$ form 
on electron time scales while the faster ones develop slower, typically during about 10-100 
inverse ion plasma frequencies. The instability, which results in the formation of shocks, 
does not set in immediately after a collision of plasma clouds at a high speed and the 
formation time of the shock may be unpredictable. We can not exclude an eventual formation
of a shock in our simulation that used the fastest collision speed. One possibility is that
the ion acoustic instability, which can develop in multidimensional systems, pre-heats the 
plasma. The Mach number of the flow is reduced by the increasing ion acoustic speed and 
a stable shock may eventually form. Another shock formation mechanism is the destabilization
of nonlinear plasma structures such as ion phase space holes that can trigger the formation 
of shocks even in very fast flows \cite{Dieckmann}.

The shock-reflected ion beam results in electron heating in all simulations. This heating is 
not caused by instabilities such as the Buneman instability that is excluded here by the low
relative flow speed between electrons and ions. It arises from the electron acceleration in 
the ambipolar electric field of the plasma density gradient. The foreshock electron temperature 
in our simulations has been about 1/3 of the electron temperature in the downstream plasma and 
it is probably independent of the upstream temperature. An electrostatic shock moves in this 
case through a medium with an ion sound speed that depends on the electron temperature in the 
post-shock plasma and on the density of the shock-reflected ion beam. The shock-reflected
ion beam, which is essential for the electron pre-heating, implies that the shock must be 
super-critical. 

The difference between the electron temperature in the foreshock and in the
upstream is moderate in our simulations and at solar system shocks. The difference can become significant at astrophysical shocks, 
where the post-shock electron temperature can be of the order of keV while that in the upstream 
is about 1 eV. A foreshock temperature of SNR shocks that is a significant fraction of the 
downstream temperature implies that the Mach number of SNR shocks is in fact reduced by at 
least an order of magnitude compared to the least upper bound $10^2-10^3$, which estimates the
ion acoustic speed using the ISM temperature $\sim 1$ eV.

The shock does not convert instantly the directed flow energy of the upstream into downstream heat. 
The shock potential and the ion distribution gradually change over a spatial interval that can be 
large if the Mach number is low, as our simulation shows. This has potentially important consequences
for experimental observations of slow shocks. Their potential jump between the downstream region and
the foreshock is moderate and it falls off over hundreds of Debye lengths. The electric field 
amplitude is thus low and smeared out over at least tens of Debye lengths. Slow shocks may thus
not always be detectable in experiments that measure the electric field amplitude. 

Finally our simulations have shown that the ions downstream have a bi-Maxwellian velocity distribution
and they are not in a thermal equilibrium. The shock-reflected ion beam is thus not the only process 
by which a supercritical shock can get rid of an excess flow energy of the upstream plasma. This aspect together with the non-Maxwellian electron velocity distributions may help explaining why electrostatic shocks are stable even if electrons and ions have the same temperature, which is not possible according to some analytic models.

\textbf{Acknowledgements:} MED wants to thank Vetenskapsr\aa det for financial support. MP acknowledges support through grant PO 1508/1-1 of the Deutsche
Forschungsgemeinschaft (DFG). LR acknowledges support from the ULIMAC grant, Triangle de la Physique RTRA. GS thanks the Leverhulme foundation (grant ECF-2011-383) for financial support. The computer time and support has been provided by the High Performance Computer Centre North (HPC2N) in Ume\aa .

\end{document}